\documentclass[journal]{IEEEtran}

\usepackage{cite}
\usepackage{amsmath,amssymb,amsfonts}
\usepackage{algorithmic}
\usepackage{graphicx}
\usepackage{textcomp}
\usepackage{booktabs}
\usepackage{multirow}
\usepackage{color, hyperref}

\hypersetup{
    colorlinks=true,
    citecolor=blue,
    linkcolor=black,
    urlcolor=blue
}

\newcommand{\argmin}{\mathop{\rm arg~min}\limits}

\def\BibTeX{{\rm B\kern-.05em{\sc i\kern-.025em b}\kern-.08em
    T\kern-.1667em\lower.7ex\hbox{E}\kern-.125emX}}
\begin{document}
\title{IGCN: Image-to-graph Convolutional Network for 2D/3D Deformable Registration}
\author{Megumi Nakao, \IEEEmembership{Member, IEEE}, Mitsuhiro Nakamura, Tetsuya Matsuda, \IEEEmembership{Member, IEEE}
\thanks{This research was supported by JSPS Grant-in-Aid for Scientific Research (B) (grant number 18H02766 and 19H04484).}
\thanks{M. Nakao and T. Matsuda are with Graduate School of Informatics, Kyoto University. Yoshida-Honmachi, Sakyo, Kyoto, 606-8501, JAPAN.}
\thanks{M. Nakamura is with Graduate School of Medicine, Kyoto University. Shogoin-Kawahara-cho, Sakyo, Kyoto, 606-8507, JAPAN.}}
\maketitle
\begin{abstract}
Organ shape reconstruction based on a single-projection image during treatment has wide clinical scope, e.g., in image-guided radiotherapy and surgical guidance. We propose an image-to-graph convolutional network that achieves deformable registration of a 3D organ mesh for a single-viewpoint 2D projection image. This framework enables simultaneous training of two types of transformation: from the 2D projection image to a displacement map, and from the sampled per-vertex feature to a 3D displacement that satisfies the geometrical constraint of the mesh structure. Assuming application to radiation therapy, the 2D/3D deformable registration performance is verified for multiple abdominal organs that have not been targeted to date, i.e., the liver, stomach, duodenum, and kidney, and for pancreatic cancer. The experimental results show shape prediction considering relationships among multiple organs can be used to predict respiratory motion and deformation from digitally reconstructed radiographs with clinically acceptable accuracy. \end{abstract}


\section{Introduction}
\IEEEPARstart{O}{rgan} positions and shapes from 3D medical images constitute patient-specific morphological information that is essential to diagnosis and pre-treatment planning. However, organs may move or deform during surgical treatment or through several weeks of radiation therapy \cite{Rigaud19, Tokuno20}. Post-imaging time-series shape changes in organs can prevent tumor localization and hinder treatment. Existing imaging devices for use during treatment have certain limitations; thus, 2D images facilitating real-time measurements (e.g., endoscopic and X-ray images) are available but 3D imaging is limited \cite{Teske15, Zhao19, Takahashi20}.

Organ shape reconstruction based on a single-projection image during treatment has wide clinical scope including image-guided therapy/intervention. However, this problem is ill-posed without prior knowledge as it requires transformation of 2D space points into points in a higher-dimensional space. One solution is 2D/3D registration, which involves the patient-specific organ shape from dense 3D computed tomography (CT) or magnetic resonance imaging (MRI) images taken prior to treatment and use of these data as prior knowledge. This approach aims to solve alignment and deformation of the organ-shape models to 2D projection images in real time, and has undergone intensive research in the field of medical image analysis over the past decade \cite{Markelj12, Reyneke19}. In particular, many studies have examined rigid-body 2D/3D image registration \cite{Wang17}\cite{Liao19} as an optimization problem for a parameter sets that determine the position and orientation.

2D/3D deformable registration of soft organs requires local point-to-point correspondence between 2D images and 3D volumes. Unlike rigid-body registration, large-scale parameters must be optimized proportional to the number of sampling points. Deformable registration between 3D volumes poses a similar problem \cite{Sotiras13}. Diffeomorphic mapping-based regularization \cite{Faisal05, Ruhaak17} enables calculation of a displacement field that can obtain smooth correspondence between sampling points; however, pairwise optimization has high computational cost for a large-scale parameter set. Thus, recent studies have investigated 3D displacement field learning using a convolutional neural network (CNN) \cite{VoxelMorph, Bob19, Krebs19, Zhao20, Tang20, Lei20}. Notably, machine learning models trained via parallel computing with a graphics processing unit can provide accelerated registration.

Single-image-based 2D/3D deformable registration has less constraints than the above-mentioned registration between 3D volumes, making stable optimization difficult. Predictions based on input images alone have high uncertainty, and the mapping between the organ shape model and 2D images, along with its learning method, are key. Some studies use bi-planer X-ray images to improve prediction accuracy\cite{Ying19, Kasten20}. In the field of computer vision, human posture and various general objects have been investigated, with a camera image database corresponding to a 3D shape being used as a background \cite{Pointnet}\cite{P2M}. Recent works have proposed integrating the CNN and a graph convolutional network (GCN) \cite{GCN}, or an estimation framework that is robust against occlusion through a self-attention network \cite{P2M}\cite{ViT}.

As regards medical imaging, collection of organ deformation data paired with 2D images is difficult, and few cases have been reported to date. A learning method for a registration map involving correspondence between an area on a 2D projection image and a local area in a 3D volume using a CNN has been proposed \cite{Miao16, Miao18, Schaffert20}. Additionally, for 2D/3D deformable registration of soft organs for surgical guidance, model-based optimization for endoscopic images has been attempted \cite{Nakao14, Saito15, Koo17, Ketcha17, Modrzejewski18}. However, within the scope of our survey, no studies have provided a framework for deep learning-based 2D/3D deformable model registration of abdominal soft organs, and no empirical cases using actual patient data have been reported.

This study introduces an image-to-graph convolutional network (IGCN) that enables 3D organ-shape reconstruction and localization based on a single-viewpoint 2D projection image. The IGCN provides a new end-to-end framework that achieves real-time 2D/3D deformable registration through integration of an image-based generative network and GCN. The generative network learns the transformation from the 2D projection image to a displacement map based on pairwise 3D meshes obtained before and after deformation. The GCN samples the input features of each node from the generated displacement field and learns the transformation into a final 3D displacement vector that satisfies the geometrical constraints. Finally, the IGCN outputs a 3D mesh, the position and deformation of which are registered to the input 2D projection image.

\begin{figure}[t]
	\begin{center}
		\includegraphics[width=8.5cm]{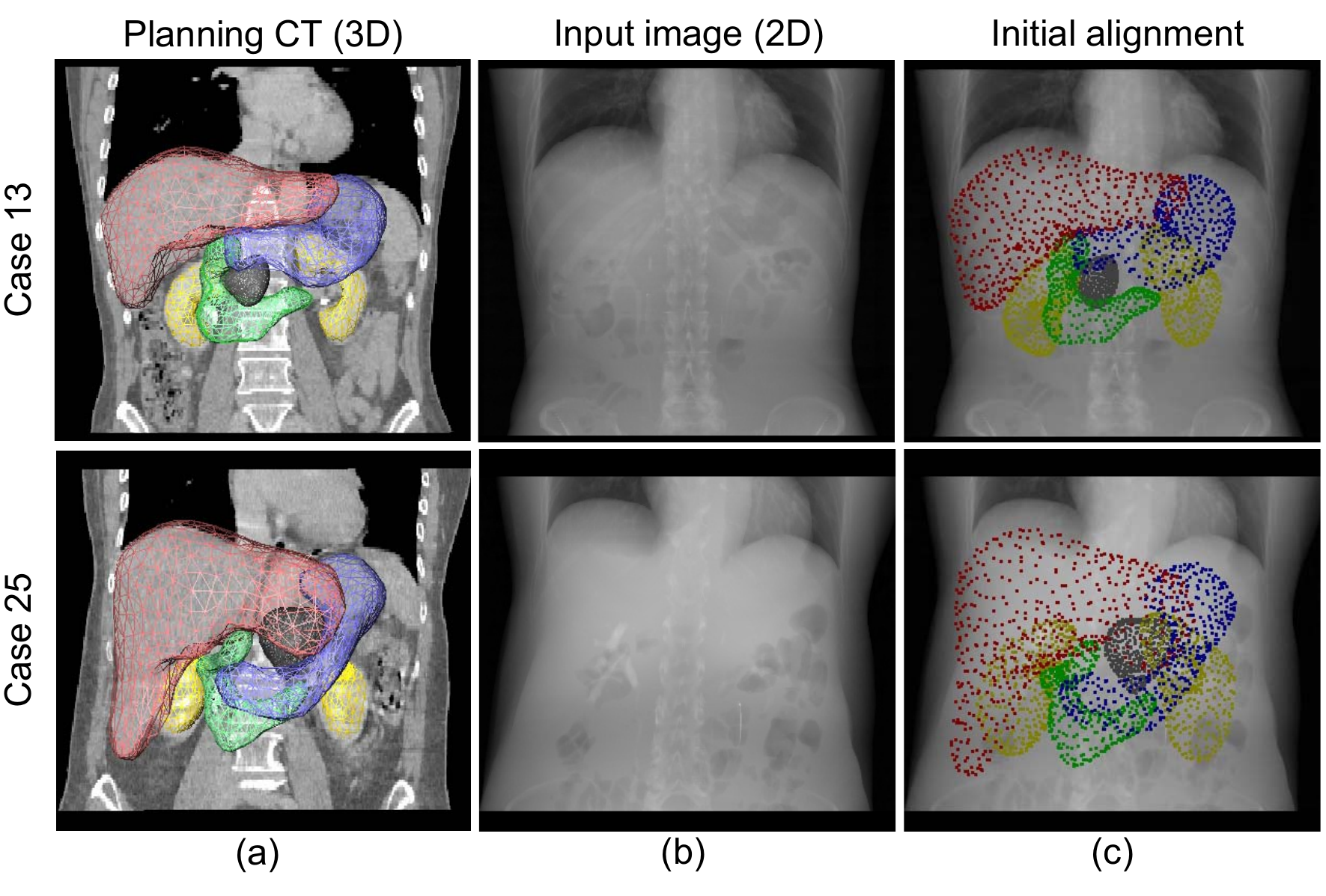}
	\end{center}
	\caption{Problem definition for 2D/3D deformable model registration for abdominal organs. (a) Pre-treatment 3D-CT and surface meshes of liver, stomach, duodenum, kidneys, and pancreatic GTV. 2D projection images (b) of target states and (c) overlaid with projected vertices of pre-treatment CT meshes.}
	\label{fig:1}
\end{figure}

Assuming application to radiation therapy, the shape reconstruction performance from a single 2D projection image targeting the abdominal organs of actual patients is verified experimentally. This is the first study to demonstrate 2D/3D deformable registration of the liver, duodenum, and kidney, and the gross tumor volume (GTV) of pancreatic cancer (Fig. \ref{fig:1}). Many variations in organ shape and deformation exist between patients, and there are almost no visual clues (such as contours) in the low-contrast 2D projection images; thus, accurate prediction of the organ positions and shapes was previously considered difficult. We also show that respiratory dynamics and deformation can be predicted from digitally reconstructed radiograph (DRR) images via statistical data augmentation for 3D-CT and simultaneous prediction of multiple organ shapes.

The methods reported herein extend a preliminary framework \cite{MICCAI21} presented at the 2021 International Conference on Medical Image Computing and Computer Assisted Intervention (MICCAI). The contributions of this study are as follows:

\begin{itemize}
    \item A new 2D/3D deformable model registration framework that integrates a displacement field generator and GCN;
    \item Simultaneous shape reconstruction of five abdominal organs, the contours of which are not directly visible on a 2D projection image;
    \item Application to localization of GTV and organ-at-risk (OAR) volumes assuming dynamic tumor-tracking radiotherapy with clinically acceptable estimation accuracy;
    \item Augmentation of respiratory deformation data based on a statistical generative model.
\end{itemize}

\section{Related Work}
\subsection{Optimization-based Approaches}
Optimization-based 2D/3D registration has been extensively researched over the past 20 years \cite{Markelj12, Brock03, Chou13, Mitrovic13, Silva16}. Rigid-body and deformable registration involve formulation as a transformation matrix or as a displacement vector field optimization problem, respectively \cite{Wang17, Mitrovic17, Lange20}. Because of the high density and large scale of 3D volumes in particular, approaches that construct shape models based on anatomical labels and seek positions and deformation on 2D images have been very successful. Notably, mesh-based shape representation can express organ elastic properties \cite{Koo17, Nakao10, Suwelack14} and statistical shape variations \cite{Rigaud19, Ehrhardt11, Nakao19} with few variables and high computational efficiency. Deformable model registration of X-ray and endoscopic images has been attempted, with the aim of surgical guidance \cite{Saito15, Koo17, Ketcha17, Modrzejewski18}. In radiation therapy, contour definitions and statistical atlases of GTVs and OARs \cite{Rigaud19, Reyneke19, Nakamura21, Nakao21} can be directly applied in model-based registration. However, 2D/3D registration based on model parameter optimization is limited to the subspace of physical models and statistical shape models defined in advance by the expressible shape variations. Additionally, the high degrees of freedom of the position and deformation in a 2D image make it difficult to set objective functions from which a stable solution can be derived, and each registration requires time for iterative optimization calculations. 

\begin{figure*}[t]
    \centering
    \includegraphics[width=15.0cm]{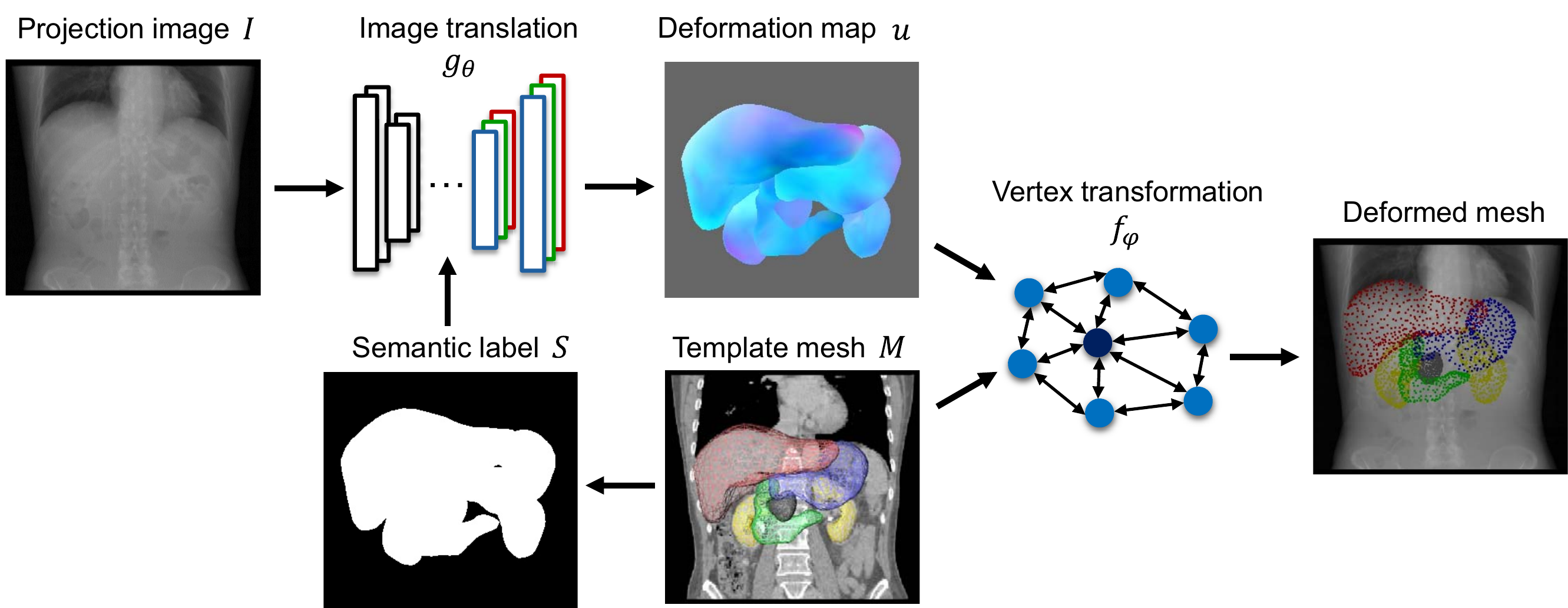}
    \caption{Full IGCN. The image translation network $g$ learns to generate a displacement field $u$. The vertex coordinates and corresponding 3D displacement vectors are concatenated into the GCN for mesh deformation learning.}
    \label{fig:2}
\end{figure*}

\subsection{Learning-based Approaches}
Recently, deep learning-based frameworks have attracted attention \cite{Toth18, Wu19, Wang20} because of the high uncertainty and computational costs inherent in model-based 2D/3D deformable registration. Pointnet \cite{Pointnet}, a CNN-based framework that generates a 3D point cloud from a single-viewpoint image, was applied to 2D virtual images of statistical pneumothorax lung models, and lobe shapes were reconstructed \cite{Wu19}. However, in point cloud representation, surface and topological information on the inter-vertex relationships, which are important for deformation field computation, are lost. Wang at al. proposed Pixel2Mesh (P2M) \cite{P2M} to generate a 3D mesh from a 2D projection image. P2M uses latent image features to deform an ellipsoid template into the target shape. A recent work \cite{Wang20} was the first to apply P2M to respiratory deformation estimation from a DRR, with 3D lung shapes being artificially generated from multiple initial 3D templates with free-form deformation. We previously implemented 2D/3D deformable registration methods using 4D-CT data for real patients \cite{MICCAI21, Tong20} and reported preliminary liver shape reconstruction results. However, in the abdominal regions, the available 2D contours or visual cues are poor. We found limitations in learning dense deformation fields and capturing distant features from low-contrast projection images. The improved IGCN framework presented herein addresses these problems and exhibits good estimation performance for multiple abdominal organs.

\section{Methods}
We consider 3D-CT/MRI volumes obtained for pre-treatment planning and unregistered 2D projection images obtained during image-guided therapy. We do not focus on automatic segmentation techniques, and we assume that the organ contours are segmented and organ shapes are modeled as triangular surface meshes as a preprocessing step.

Let $M$ be the initial mesh generated from 3D-CT planning images and $I$ be the 2D projection image obtained from the target state. X-ray or endoscopic images are candidates for $I$; however, in our experiments for quantitative evaluation of the proposed 2D/3D deformable registration method, DRR images were used. These DRR images were generated from 4D-CT images for performance analysis targeting non-linear motion and shape variability of abdominal organs during respiratory motion.

The left and central images of Fig. \ref{fig:1} show two typical examples of $M$ and $I$, respectively. In the projection images, the abdominal organs are invisible. Further, the anatomical variability in organ shape and location between patients is apparent. The right images are overlaid with the projected vertices of the mesh, indicating initial misalignment between $M$ and $I$. The diaphragm shape visualized in the DRR does not match the projected initial shape because the two states differ in terms of patient condition (e.g., posture and respiratory phase). The deformation is nonlinear and exhibits local rotation and sliding motion \cite{Nakao21} in 3D, and simple linear transformation is not sufficient to register the two states.

\subsection{IGCN Architecture}
Fig. \ref{fig:2} is an outline of the IGCN architecture and deformable model registration process. The IGCN is a generalized, organ-independent framework that integrates an image translation network $g$ and a vertex transformation network $f$. Various architectures are acceptable for each network model; we employed a UNet-based network \cite{Unet} and graph convolutional network \cite{GCN} for $g$ and $f$, respectively. 

$g$ takes a 2-channel image formed by $I$ and a semantic label $S$ generated from $M$. In our experiments, the input image size was $640 \times 640 \times 2$; however, the method is not limited to a particular size. $g$ learns the generation of a displacement map $u$, which represents a spatial mapping function in 2D space. Then, $f$ receives feature vectors from $u$ and $M$. $M$ is projected onto the 2D image space, and the pixel values of $u$ that correspond to each template vertex are sampled. The 3D vertex coordinates of $M$ and corresponding 3D displacement vectors are concatenated into $f$ for learning deformation. Finally, $f$ generates a deformed mesh registered to $I$.

The IGCN implements a new 2D/3D deformation learning scheme characterized by $f$ and $g$. Existing projected point sampling methods struggle to capture image features distant from the initial mesh\cite{P2M, MICCAI21}. P2M \cite{P2M} employs CNN-based feature encoding with hierarchical extension to fit an ellipsoid template to various 3D objects. However, it concentrates on mesh deformation and neglects the target object motion. Abdominal organs contain both local deformation and global translation, and the projection images have no clear edges. The displacement map $u$ and composite function $g \circ f$ address the non-linear organ motion and deformation. We describe the roles of these two functions in the next sections.

\subsection{Displacement Mapping Function}
CNN-based encoding schemes struggle to learn image features distant from the initial template. In our preliminary study \cite{MICCAI21}, we mapped the projection point to a new position for which a higher probability of obtaining effective image features was expected. Here, this method is referred as ``IGCN Warp.'' This scheme improved the registration results; however, when the deep features transformed from the input image were discretely referenced at the sparse vertex level of the mesh, the deep feature creation in the CNN and the pairwise updating of the mesh vertex coordinates were likely to become unstable.

\begin{figure}[t]
    \centering
    \includegraphics[width=8.5cm]{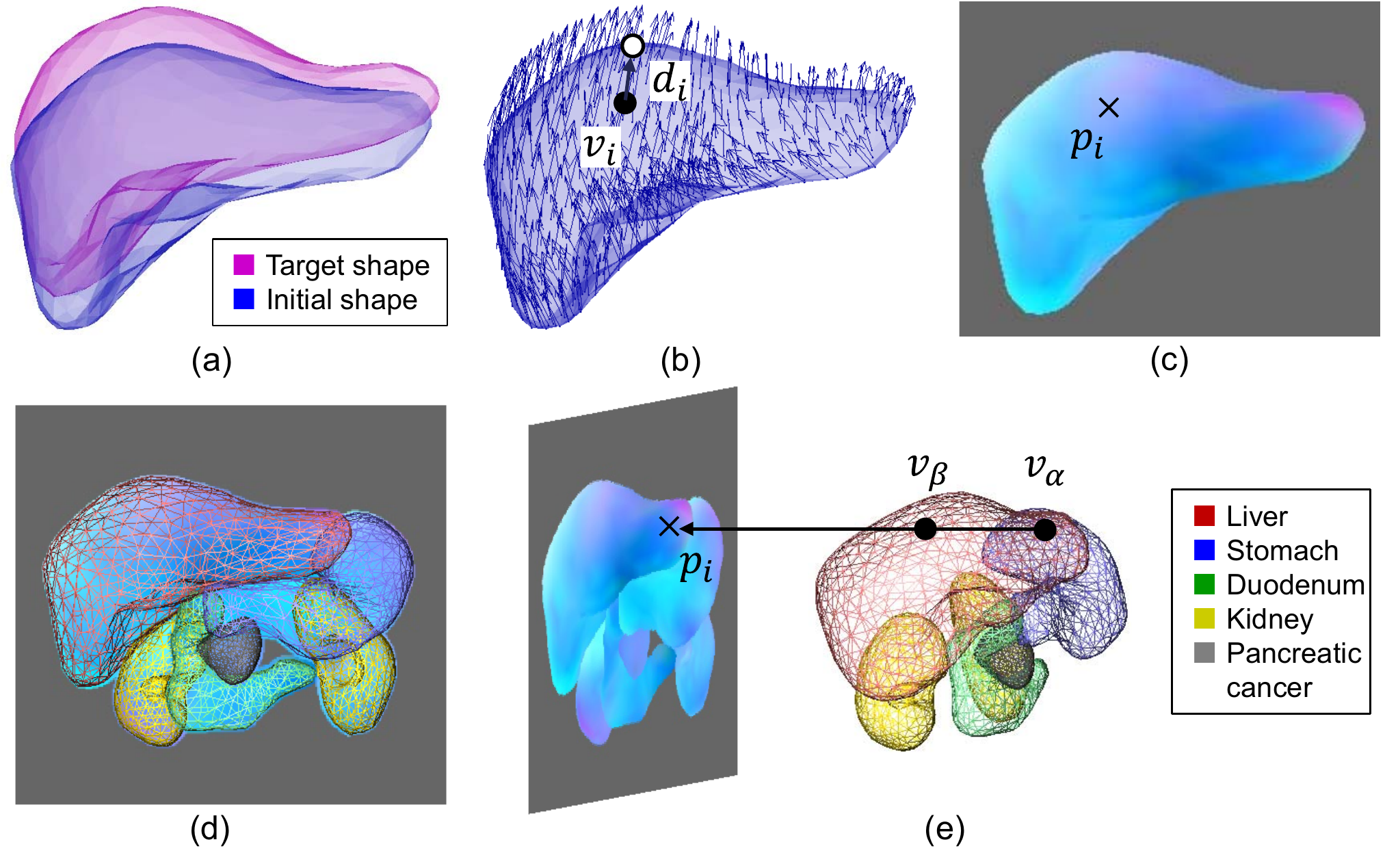}
    \caption{Learning scheme for image translation function generating spatial mapping function $u$.  (a) DMR between initial and target shapes, (b) 3D displacement vectors obtained from corresponding vertices, (c) forward displacement map and sampling, (d) displacement map in abdominal regions overlaid with their meshes, and (e) one-to-many correspondence between displacement field and mesh.}
\label{fig:3}
\end{figure}

The method newly proposed in this study involves learning of the transformation function $g_\theta$ from $I$ to $u$ based on the supervised learning framework via the image generative network. Fig. \ref{fig:3} illustrates the $g_\theta$ learning process for the liver. First, we obtain a registered mesh with point-to-point correspondence through deformable mesh registration (DMR) \cite{Nakao21} between the initial and target meshes (Fig. \ref{fig:3}(a)). The displacement vector $d_i$ of each vertex $v_i$ is obtained from the corresponding points before and after deformation (Fig. \ref{fig:3}(b)). A 3-channel projection image (Fig. \ref{fig:3}(c)) is obtained by transforming $d_i$ from Euclidean to color space and is then used as the surface color of the initial mesh for rendering. This is a forward displacement map in which a 3D displacement vector is stored in each pixel, which directly represents $u$.

The 2D region of the patient-specific organ obtained by projecting the initial mesh can be used as a semantic attention label $S$. Here, the proposed $g_\theta$ defines the transformation:
\begin{eqnarray}
    u = g_\theta(I, S).
\end{eqnarray}
$S$ is used as additional attribute information, and both $I$ and $S$ are treated as the 2-channel input image for stable learning and convergence of the network parameters.

Fig. \ref{fig:3}(d) shows the $u$ used for supervised learning and formed from the meshes of the five abdominal organs. Here, $u$ is a projection of the volumetric displacement field that expresses the 3D mesh deformation; thus, the projection points $p_i$ in the map are referenced from the multiple vertices $v_\alpha, v_\beta$ in the mesh (Fig. \ref{fig:3}(e)). In this case, identical displacement vectors can be assigned to all vertices mapped to $p$. However, $v_\alpha, v_\beta$ form parts of different organs and different parts of the same organ (e.g., the anterior and posterior); thus, they must each be able to express different displacements. This problem is resolved in the GCN described below through embedded learning using 3D vectors obtained from the displacement map as well as the local shape and topologies at each mesh vertex.

\subsection{Vertex Transformation Function}
The vertex transformation function $f$ updates each vertex in the mesh using the generated $u$ and template mesh $M$ structure. Thus, $f$ is responsible for the spatial transformation of each vertex in the mesh based on the GCN, where 
\begin{eqnarray}
    \hat{v_i} = f_\varphi(v_i, u(p_i)).
\end{eqnarray}
Here, $v_i$ is the vertex coordinates after normalization; $u(pi)$ is the 3D displacement vector obtained from the corresponding projection point $p_i$ in the displacement map; $f_\varphi$ is composed of the GCN and a learnable parameter $\varphi$, where the input is a vector having $v_i$ and $u(p_i)$ concatenated; and the output is the predicted value $\hat{v_i}$ of the vertex coordinates. Deformation of the entire mesh is calculated by transforming all vertices $v_{i} \in \mathcal{V} (i = 1,2,$\ldots$,n)$ composing $M$ using the trained function $f_{\hat{\varphi}}$.

For the GCN layers, graph convolution is applied to obtain hierarchical topological features in non-Euclidean space \cite{GCN}. The mesh is a type of graph ${G(\mathcal{V}, \mathcal{E})}$, where $\mathcal{V}$ is the set of vertices and $\mathcal{E}$ is the set of edges. The per-vertex features are shared with the neighbor vertices. The GCN employed in this study consisted of eight sequential graph convolutional layers, each of which is defined in Eq. (3).
\begin{eqnarray}
    X^{(l+1)}=\sigma(\hat D^-{}^\frac{1}{2}\hat A \hat D^-{}^\frac{1}{2} X^{(l)} W^{(l)}) ,
\end{eqnarray}   
where $X^{(l)}$ and $X^{(l+1)}$ denote the feature matrix before and after convolution, respectively. In our experiments, $X^{(l)}$ was the concatenation of the vertex coordinates $v_i$ and displacement vectors $u(p_i)$, and $W$ was the learnable parameter matrix. ${A}\in \mathbb{R}^{n \times n}$ was the adjacency matrix, i.e., a symmetric matrix with binary values, in which element $A_{ij}$ was 1 if there was an edge between $v_i$ and $v_j$, and 0 if the two vertices were not connected. ${D}\in \mathbb{R}^{n \times n}$ was the degree matrix, i.e., a diagonal matrix, in which each element $A_{ii}$ represented the number of edges connected to $v_i$. The template mesh was deformed by updating $X^{(l)}$.

\subsection{Loss Functions}
The parameters $(\theta, \varphi)$ of the overall network are simultaneously updated and optimized by minimizing an objective function. In this section, we introduce three loss functions to achieve 2D/3D deformable mesh registration under the constraint of smooth deformation.

The ground-truth vertex coordinates of the target meshes are obtained from the deformable registration process. To strictly evaluate the point-to-point correspondence, we define the mean distance loss $\mathcal{L}_{pos}$ of the vertex positions between the estimated shape and the ground truth as
\begin{eqnarray}
    \mathcal{L}_{pos}=\frac{1}{n} \sum_{i=1}^{n}\|v_{i}-\hat{v}_{i}\|^{2}_{2} ,
\end{eqnarray}
where $v_{i} \in \mathcal{V} (i = 1,2,$\ldots$,n)$ is the target 3D position, and $\hat{v}_{i}$ is the predicted position. This loss function induces the convergence of the estimated vertex to the correct position.

In our problem setting, the organ deformation is spatially non-linear and heterogeneous but expected to remain within a limited range. To preserve the curvature and smoothness of the initial surface, we use a regularization loss $\mathcal{L}_{smooth}$ that evaluates a discrete Laplacian of the mesh: 
\begin{eqnarray}
    \mathcal{L}_{smooth} = \frac{1}{n}\sum_{i=0}^{n}\|L(v_{i})-L(\hat{v}_{i})\|^{2}_{2} ,
\end{eqnarray}
where $L (\cdot)$ is the Laplace-Beltrami operator and $L(v_i)$ is the discrete Laplacian of $v_{i}$ defined by $L(v_i) = \sum_{j \in N(v_i)} (v_i -v_j)/N(v_i)$. Here, $N(v_i)$ is the number of adjacent vertices $v_j$ of the 1-ring connected by the vertex $v_i$. This loss constrains the shape changes from the initial state and avoids generation of unexpected surface noise and low-quality meshes.

In addition to evaluating the mesh vertex coordinates, accurate prediction of $u$ improves the 2D/3D deformable registration results. Specifically, stable learning of $u$ is important when the target contains both translation and local deformation. Thus, we introduce the displacement map loss $\mathcal{L}_{map}$ determined by the mean absolute error (MAE), such that
\begin{eqnarray}
       \mathcal{L}_{map}= \|u - \hat{u}\|_{1} ,
\end{eqnarray}
where $u$ is the target displacement map and $\hat{u} = g(I, S)$ is the predicted displacement map translated from $I$. The existing 2D/3D deformable registration framework \cite{P2M, Wang20} does not use a displacement map, and this study is the first to investigate the performance of the newly designed loss function.

The full objective $\mathcal{L}$ is defined as the weighted sum of the above three loss functions:
\begin{eqnarray}
    \mathcal{L} = \mathcal{L}_{pos}+\mu\mathcal{L}_{map}+\lambda\mathcal{L}_{smooth}.
\end{eqnarray}
The loss function values are normalized to $[0, 1]$ using the maximum values in each space. Here, to facilitate supervised learning, we used 1.0 and 0.1 for $\mu$ and $\lambda$, respectively, after examination of several parameter sets. The optimized deformable registration model $g_{\theta^*} \circ f_{\varphi^*}$ is obtained by solving
\begin{eqnarray}
    g_{\theta^*}, f_{\varphi^*} = \argmin_{g_{\theta}, f_{\varphi}} \mathcal{L} (g_{\theta}, f_{\varphi}).
\end{eqnarray}
These are applied to the developed framework at each epoch to train $(g_{\theta}, f_{\varphi})$.

\begin{figure}[t]
    \centering
    \includegraphics[width=8.5cm]{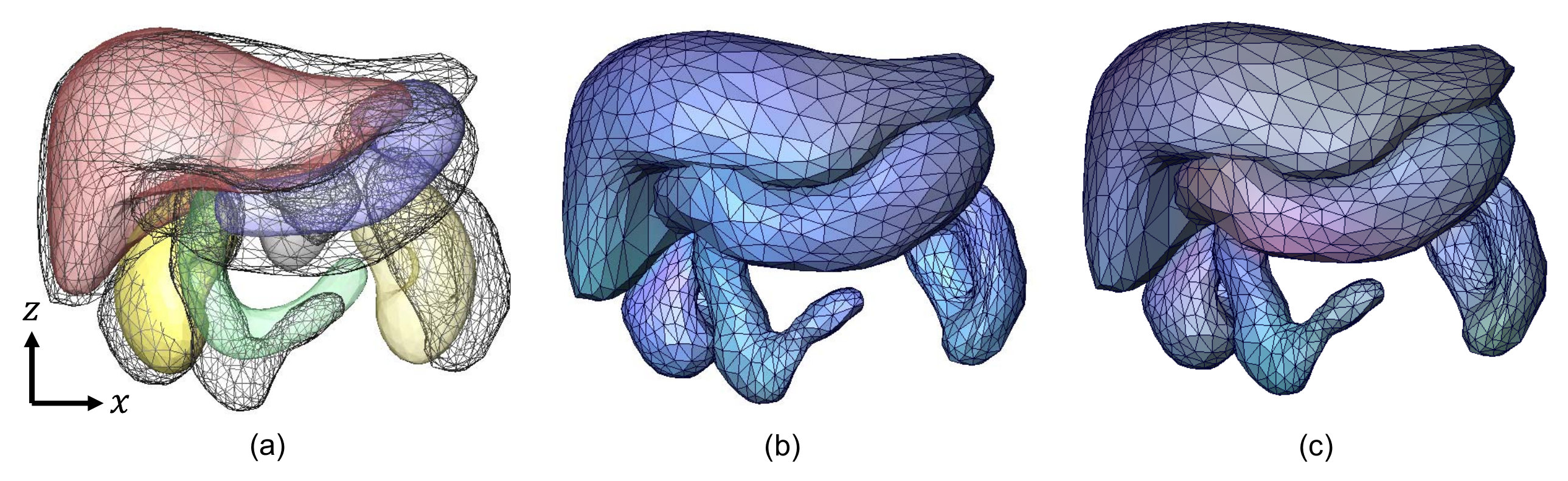}
    \caption{Statistical models of abdominal organs with respiratory deformation: (a) mean (translucent) and patient-specific (mesh) shapes, and (b) first and (c) second principal components of vertex displacements. The colors represent 3D displacement vectors.}
\label{fig:4}
\end{figure}

\subsection{Statistical Generative Models}
In this section, we introduce a data augmentation method based on statistical generative models to overcome the limited training data volumes. Displacements that reflect the statistical properties of respiratory deformation, as obtained from 4D-CT data, are supplied to a mesh obtained from 3D-CT data. Specifically, for a mesh generated from a 4D-CT volume, DMR \cite{Nakao21} is conducted between all cases to obtain a mesh with the same topology. Then, principal component analysis is conducted to obtain a statistical model of the shape and displacement. 

Fig. \ref{fig:4}(a) shows patient-specific shapes obtained via 4D-CT and the average shapes calculated from data for multiple patients. Figs. 4(b) and (c) show results obtained by transforming the first and second principal components, respectively, of the displacement to the RGB space and visualizing these data as color maps for the mesh surfaces. The displacement z-component is large because of the characteristics of respiratory motion; however, the local displacement distributions of each organ have different orientations and sizes. The statistical $d_i$ is defined as 
\begin{eqnarray}
    d_i = \sum_{k=0}^{m} \omega_k c_i^{(k)},
\end{eqnarray}
where $c_i^{(0)}$ is the mean displacement at vertex $v_i$ and $c_i^{(k)} (k \geq 1)$ is the $k$th component of the displacement. Further, $\omega_k$  is the weight parameter for each component, and can be changed to yield various $d_i$ values and express the statistical deformation of the 4D-CT data.

Augmented data for supervised learning can be obtained by deforming the registered mesh $M$ obtained from the 3D-CT volume based on $d_i$. In other words, the vertex coordinates are updated for each $v_i$ of $M$ as $v_i \leftarrow v_i - d_i$. The set of the deformed mesh $M_d$ and the projection image $I$ obtained from 3D-CT volume is used as the input data. The pre-update mesh $M$ can be used as the target shape of the true value corresponding to $I$. Network $g \circ f$ training is implemented by randomly changing $\omega_k$ for each epoch and generating augmented data with various deformation variations online. 

\section{Experiments}
To verify the performance of the proposed method and its potential application in clinical settings, we conducted the following two experiments: comparison with conventional methods of 2D/3D deformable registration and deformation prediction for multiple organs, assuming clinical applications to moving-target tracking radiation therapy. We implemented our methods using Python 3.9 and TFlearn with a TensorFlow background. We used 1 for each training batch, 300 for the total number of training epochs, and the ADAM optimizer with a learning rate of $1\times 10^{-4}$. Our code and demonstration movies are available online at \href{https://github.com/meguminakao/IGCN}{https://github.com/meguminakao/IGCN}.

\subsection{Dataset}
3D-CT volumes of 124 cases and 4D-CT volumes of 35 cases were acquired from various patients who underwent intensity-modulated radiotherapy in Kyoto University Hospital. This study was performed in accordance with the Declaration of Helsinki and was approved by our institutional review board (approval number: R1446). Each 4D-CT volume consisted of 10 time phases ($t$=0, 10, $\cdots$, 90$\%$) for one respiratory cycle and was measured under respiratory synchronization, with $t$ = 0 and $t$ = 50 corresponding to the end-inhalation and end-exhalation phases, respectively. Thus, 474 3D-CT volumes were used.

Each 3D-CT volume consisted of 512 $\times$ 512 pixels and 88-152 slices (voxel resolution: 1.0\text{~mm} $\times$ 1.0\text{~mm} $\times$ 2.5\text{~mm}). During routine clinical procedures, the following regions were labeled by board-certified radiation oncologists: the entire body, stomach, liver, duodenum, left and right kidneys, and the clinical target volume (CTV). We generated surface meshes (400-500 vertices and 796-996 triangles for one organ) from the region labels and obtained organ mesh models with point-to-point correspondence using DMR. The DMR algorithm and the registration performance for the abdominal organ shapes were reported previously \cite{Nakao21}, and template meshes registered to patient-specific organ shapes with a 0.2\text{mm} mean distance (MD) error and 1.1\text{mm} Hausdorff distance (HD) error, on average, were confirmed. This registration error was sufficiently small for the use of ground-truth meshes.

\begin{figure*}[t]

	\begin{center}
		\includegraphics[width=17.5cm]{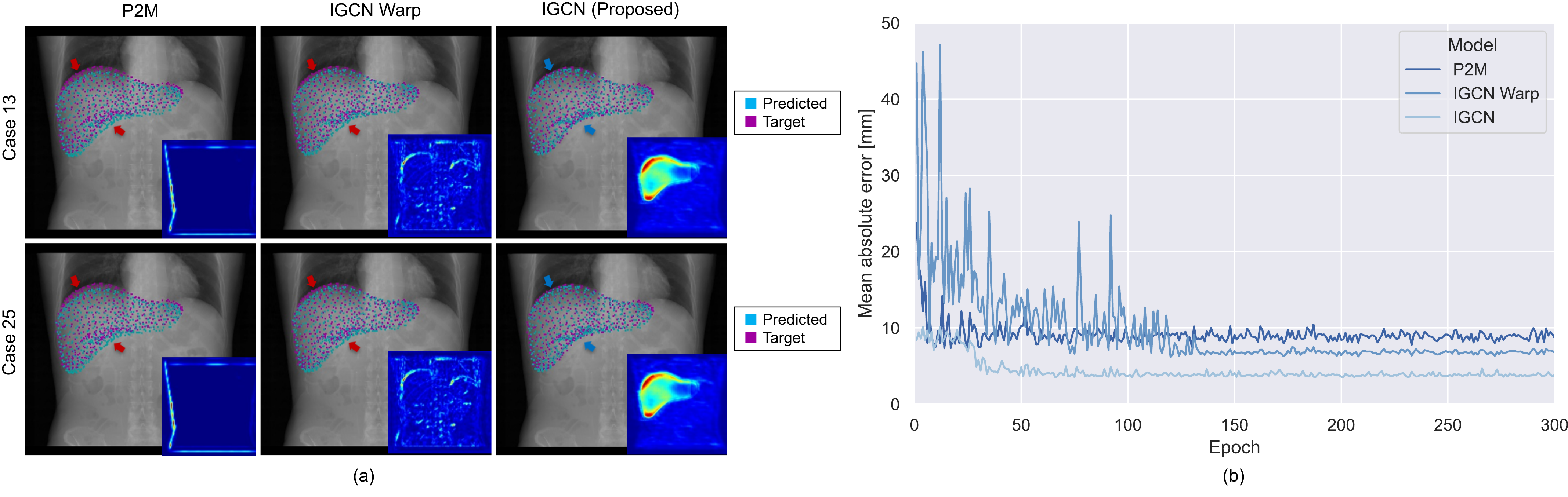}
	\end{center}
	\caption{Visual comparison of methods with respect to liver shape reconstruction, (a) registered shapes and latent image features for average (Case 13) and maximum (Case 25) error cases, and (b) learning curves.}
	\label{fig:6}
\end{figure*}

\subsection{Method Comparison}
The first experiment was designed with the aim of quantitatively and qualitatively comparing the registration results of the proposed and existing methods. The 4D-CT data were used as the test data, and the registration errors of the organ shape meshes calculated for the generated DRRs were compared. The liver was in contact with the diaphragm and the upper two-dimensional contour was detectable, but the contours on the lateral and lower regions could not be visually confirmed. The contours of the other abdominal organs (stomach, duodenum, kidney, and pancreatic cancer GTV) could not be visually confirmed on the 2D projection image, and 3D shape reconstruction was even more challenging. In this experiment, the liver was used as the estimation target to facilitate performance evaluation and error analysis.

\subsubsection{Baseline and experimental conditions}
Few existing methods can achieve 2D/3D deformable model registration from a single-viewpoint projection image for deformable organs. We selected P2M \cite{P2M}, which was proposed for general images, and IGCN Warp \cite{MICCAI21}, which we developed in previous research, as baselines for comparison, and compared their performance with that of the proposed IGCN. The true value was obtained for each vertex; thus, the Chamfer loss used in P2M was changed to the $L_{pos}$ defined in Eq. (4), and the remaining P2M loss was used without alteration. Hierarchical learning was not applied to match the prediction process with each method.

Verifications were conducted using the following two conditions with respect to initial alignment of the liver mesh: the noise-free condition, which used the position at $t$ = 0 in the first phase and corresponded to the 4D-CT end-inhalation phase; and the noisy condition, for which the initial position was set as the position translated in 3D using random noise, with the maximum displacement being twice the average respiratory displacement. We considered this noisy condition because of the difference in setup of input images between the experiment and clinical situation. For both conditions, the liver shape and position were set as unknown for all phases, dynamic properties and hysteresis caused by time changes were neglected, and this problem was regarded as a problem of static reconstruction of the liver shape in each phase. For each method, training was performed through data augmentation using the statistical generative model mentioned above.

The number of 4D-CT cases were limited; therefore, we adopted a 3-fold cross validation method, which divided 35 patients into three groups of 12, 12, and 11. We calculated the mean and Eigen displacement from 4D-CT data for 23 patients, excluding the test data; these were then adopted to the 3D-CT data of 124 patients for learning while continuously generating variations of the organ displacement associated with respiration. The weight parameters were determined as $(\omega_0, \omega_1, \omega_2) = (2, 1, 0)$ after examination of the prediction performance with several parameter sets. Regarding the selection method and the effect of the data augmentation on the prediction performance, refer to the supplementary documents.

We evaluated the 3D shape and position accuracies for the predicted organs using three error indices, mean distance (MD), Hausdorff distance (HD)\cite{Hausdorff} and mean absolute error (MAE) between surfaces, as well as the Dice similarity coefficient (DSC). We obtained a mesh with vertex correspondence by applying the DMR for each organ \cite{Nakao21} and used it as the target shape with ground-truth coordinates. The MD and HD were the average and maximum values, respectively, of the bidirectional distance defined by the two nearest vertices between the predicted and true-value mesh; these values quantified the error between shapes. The MAE was the average Euclidean distance between the predicted and correct positions of the corresponding vertex and reflected the prediction error for each vertex. The DSC quantified the overlap between the 3D regions of two meshes; a higher value indicated better prediction performance.
	
\begin{table}[t]
	\caption{Quantitative comparison of liver shape reconstruction. Mean $\pm$ standard deviation of MD, HD, MAE and DSC.}
	\label{table:1}
    \centering
    \scalebox{0.9}{
        \begin{tabular}{ccccc}
    		\hline
    	    \multirow{2}{*}{w/o noise} & \multicolumn{4}{c}{Methods}    \\ \cline{2-5} 
    	    & Initial & P2M & IGCN Warp & Proposed \\ 
    		\hline
    		MD [mm] & 3.75 $\pm$ 2.31 & 3.71 $\pm$ 1.37 & 2.93 $\pm$ 1.43 & 2.10 $\pm$ 0.83 
    		\\
    	    HD [mm] & 14.38 $\pm$ 6.77 & 12.77 $\pm$ 3.88 & 12.27 $\pm$ 5.35 & 9.90 $\pm$ 4.05 
    	    \\
    		MAE [mm] & 8.93 $\pm$ 4.68 & 8.57 $\pm$ 2.57 & 7.47 $\pm$ 3.30 & 6.08 $\pm$ 2.40  
    		\\
    		DSC [\%] & 89.45 $\pm$ 6.67 & 89.68 $\pm$ 4.11 & 91.88 $\pm$ 4.10 & 94.53 $\pm$ 2.24 
    		\\
    		\hline
        \end{tabular}
    }
    
    \vspace{5mm}

	\caption{Liver shape reconstruction results for case of random noise added to the initial alignment.}
	\label{table:2}
    \centering

    \scalebox{0.9}{
        \begin{tabular}{ccccc}
    		\hline
    		\multirow{2}{*}{w/noise} & \multicolumn{4}{c}{Methods}    \\ \cline{2-5} 
    	    & Initial & P2M & IGCN Warp & Proposed \\ 
    		\hline
    		MD [mm] & 4.52 $\pm$ 2.68 & 4.25 $\pm$ 2.25 & 3.37 $\pm$ 1.77 & 2.29 $\pm$ 0.97 
    		\\
    	    HD [mm] & 16.12 $\pm$ 7.82 & 14.24 $\pm$ 5.46 & 13.36 $\pm$ 5.93 & 10.44 $\pm$ 4.33 
    	    \\
    		MAE [mm] & 10.26 $\pm$ 5.31 & 9.61 $\pm$ 3.98 & 8.25 $\pm$ 3.72 & 6.41 $\pm$ 2.52  
    		\\
    		DSC [\%] & 86.97 $\pm$ 8.03 & 88.13 $\pm$ 6.60 & 90.51 $\pm$ 5.23 & 93.93 $\pm$ 2.71 
    		\\
    		\hline
        \end{tabular}
    }

\end{table}

\subsubsection{Comparison of results with baseline}
Table \ref{table:1} lists average values and standard deviations of the evaluation indices obtained for 350 test data points. Here, ``Initial'' refers to the magnitude of the deviation from the known 3D shape of the first phase $t$ = 0 and corresponds to the stage when deformation prediction was not performed. The proposed method exhibited superior performance to P2M and IGCN Warp, and a 3D liver shape was reconstructed with shape error values of MD = 2.1\text{~mm} and HD = 9.9\text{~mm}, and a shape similarity of DSC = 94.5$\%$. Significant differences (one-way analysis of variance, ANOVA; p $<$ 0.05) were confirmed for the conventional methods (P2M and IGCN Warp) for all indices. 

Table \ref{table:2} lists the respective errors when noise was added to the initial template alignment. For the MD values, the errors increased by 14.6$\%$ (0.5\text{~mm}) and 15.0$\%$ (0.4\text{~mm}) for P2M and IGCN Warp, respectively, but the increased error of the proposed method was suppressed to 9.0$\%$ (0.2\text{~mm}). Thus, stable prediction could be achieved even for differences in the initial conditions associated with the 3D shape arrangement. As with the noise-less conditions, significant differences were confirmed from the conventional method for all indices.

The smoothness of the predicted shape and the mesh quality could not be evaluated using the above error indices only; therefore, we qualitatively confirmed the estimation results by visualizing the estimated shape. Fig. \ref{fig:6}(a) shows results obtained by superimposing the liver shape predicted through DRR of the end-exhalation phase ($t$ = 50) for each method, for the case in which predictions were based on the mean shape error (Case 13) and the case for which the shape error was largest (Case 25). Magenta indicates the true liver shape and position, and cyan shows the predicted liver shape. A heat map on the right-bottom of each figure show the sum of the latent image features in the same feature encoding layer. 

The proposed method successfully predicted deformation similar to the target 3D shape despite the fact that visual confirmation of the contours was not achieved for many liver areas, and with only extremely low-contrast textures being confirmed. Visual comparisons of the latent image features and prediction results revealed that P2M responded strongly to the body contour edge, with large errors at locations with low correlation with the body contour movements, as indicated by the arrows in Fig. \ref{fig:6}(a). Cases in which the prediction may fail with large displacement, even if the edge around the diaphragm is relatively clear, are shown. For IGCN Warp, responses to the low-contrast edges and texture were apparent. However, the errors increased in the lower liver, where the edge could not be visually confirmed. The proposed method generated features for the liver area and its surroundings, which yielded favorable predictions for the lower area of the liver. Fig. \ref{fig:6}(b) depicts the learning curves of the three methods for MAE of the test datasets. Each method converged before 150 epochs, with IGCN converging fastest and the other two methods showing unstable curves.

\begin{figure*}[t]
	\begin{center}
		\includegraphics[width=17.5cm]{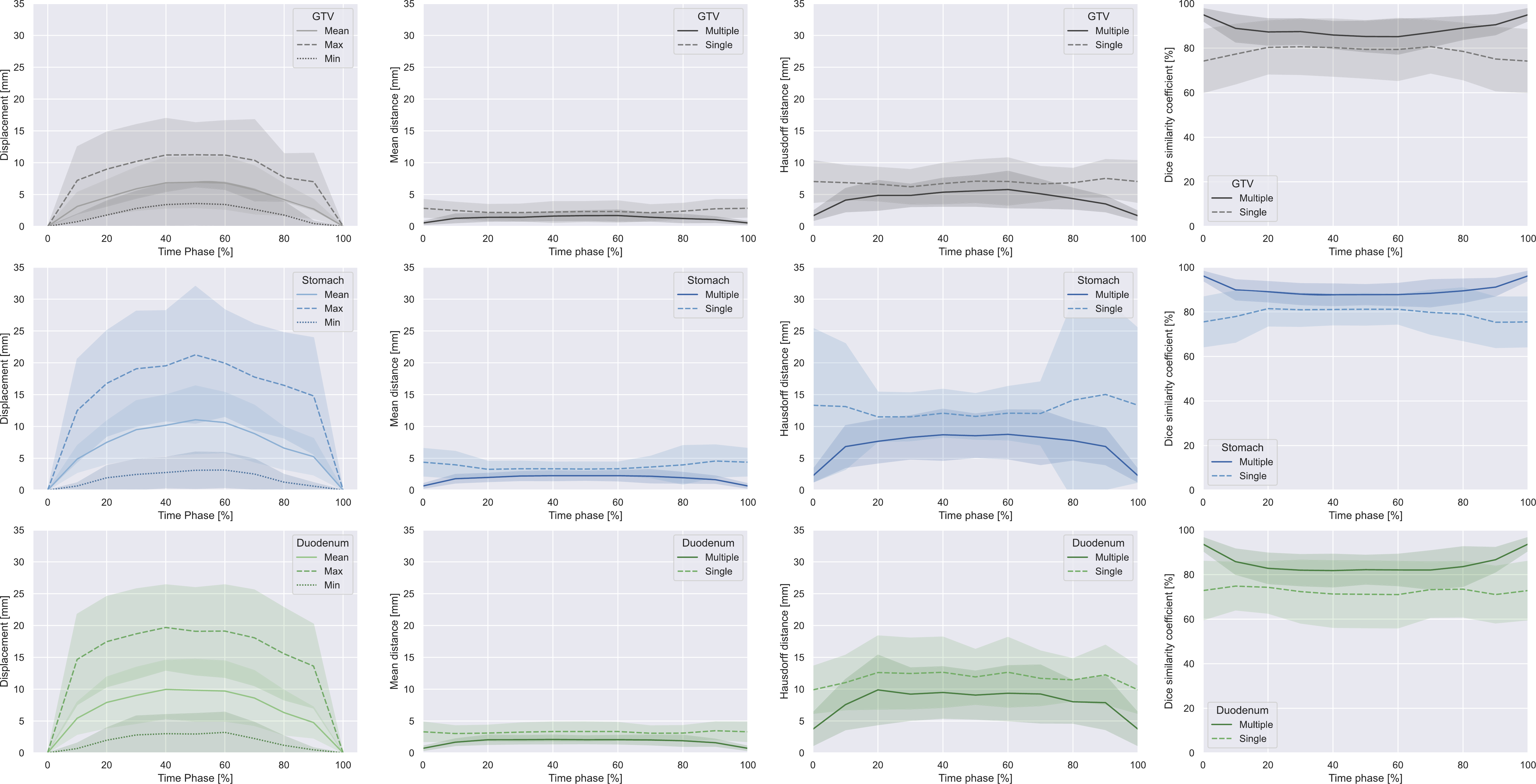}
	\end{center}
	\caption{Motion dynamics and shape reconstruction errors for three abdominal organs and pancreatic cancer GTV. The means and standard deviations of the corresponding vertices are plotted in the graphs.}
	\label{fig:7}
\end{figure*}

\subsection{Multiple-Organ Deformation Prediction}
In the final experiment, we aimed to verify the organ deformation and displacement prediction performance assuming clinical applications to moving-target tracking radiation therapy. We generated a 10-frame sequential DRRs from 4D-CT data and conducted an experiment to predict the 3D shapes of the liver, stomach, duodenum, and pancreatic cancer GTV. Two approaches can be employed to estimate the shapes of multiple organs, here referred to as ``single (SR)''and ``multiple reconstruction (MR)'': SR learns each individual organ and MR simultaneously learns multiple organs as a tetrahedral mesh by generating connectivity between organs. MR increases both the number of vertices in the mesh to be estimated and the shape expression complexity, but it may effectively learn positional relationships between organs and the deformation interactions. The estimation performance was compared for both methods. We verified whether the final performance achieved the 3D organ area identification accuracy required for adaptive radiation therapy.

As superior shape reconstruction performance can be expected when 4D-CT data are also used for training, we adopted the 3-fold cross validation method for this training Thus, 35 patients were divided into three groups of 12, 12, and 11, as in the previous section. These data were then added to the statistical generative model obtained by deforming the 3D-CT data of 124 different patients. We then conducted learning using a total of 354 volumes from the 4D-CT data (230 3D-CT volumes), for 23 patients in the remaining two groups that were not incorporated in the test data. For SR, we calculated the error by predicting each 3D shape from one DRR for the liver, stomach, duodenum, and pancreatic cancer GTV based on the trained network. For MR, we calculated the shape error for each organ after simultaneously predicting the 3D shapes of all four organs from one DRR.

\begin{table*}[t]
	\caption{Comparison of deformable registration performance for single and multiple abdominal organs.}
	\label{table:3}
  \centering
  \scalebox{0.9}{
	\begin{tabular}{cccccccc}
		\hline
		\multirow{2}{*}{} 
		         & \multicolumn{3}{c}{Single reconstruction (SR)} & & \multicolumn{3}{c}{Multiple reconstruction (MR)}\\ 
		\cline{2-4} \cline{6-8} 
	             & MD [mm] & HD [mm] & DSC [\%] & & MD [mm] & HD [mm] & DSC [\%] \\ 
		\hline
		Liver & 1.83 $\pm$ 0.89 & 8.49 $\pm$ 4.57 & 95.32 $\pm$ 2.31 & & 1.86 $\pm$ 0.89 & 8.35 $\pm$ 4.52 & 95.13 $\pm$ 2.31 \\
		Stomach & 3.59 $\pm$ 1.92 & 11.59 $\pm$ 9.24 & 80.14 $\pm$ 9.77 & & 1.77 $\pm$ 0.92 & 6.49 $\pm$ 3.93  & 90.89 $\pm$ 5.40 \\
	    Duodenum & 2.97 $\pm$ 1.38 & 11.33 $\pm$ 4.72 & 75.83 $\pm$ 13.19 & & 1.64 $\pm$ 0.82 & 7.34 $\pm$ 4.47 & 86.75 $\pm$ 7.82 \\
		GTV & 2.08 $\pm$ 1.43 & 6.42 $\pm$ 3.00 & 82.16 $\pm$ 13.25 & & 1.10 $\pm$ 0.74 & 3.86 $\pm$ 2.36 & 89.90 $\pm$ 6.56 \\
		\hline
    \end{tabular}
    }
\end{table*}

\subsubsection{Performance analysis and motion dynamics}	
Fig. \ref{fig:7} shows the liver, stomach, duodenum, and GTV displacements for each phase, as well as the shape reconstruction errors due to SR and MR. The mean displacement for all corresponding vertices was visualized as the centerline, and the standard deviation was depicted as a colored band. Table \ref{table:3} lists the errors for each organ for both SR and MR. For the liver, no significant differences between the two approaches were apparent, but significant differences (ANOVA; p $<$ 0.05) were confirmed between the two methods for the stomach, duodenum, and GTV. Shape error improvements of 50.7$\%$, 44.8$\%$, and 47.2$\%$, respectively, were obtained for MR. In the American Association of Physicists in Medicine guideline for image registration and fusion \cite{Brock16}, the quantitative metric tolerance is MD = 2 - 3\text{~mm} and DSC = 80 - 90$\%$. The obtained results show that shape reconstruction can be achieved with accuracy equal to or exceeding this level and, thus, the proposed method is clinically applicable.

\begin{figure*}[t]
	\begin{center}
		\includegraphics[width=17.5cm]{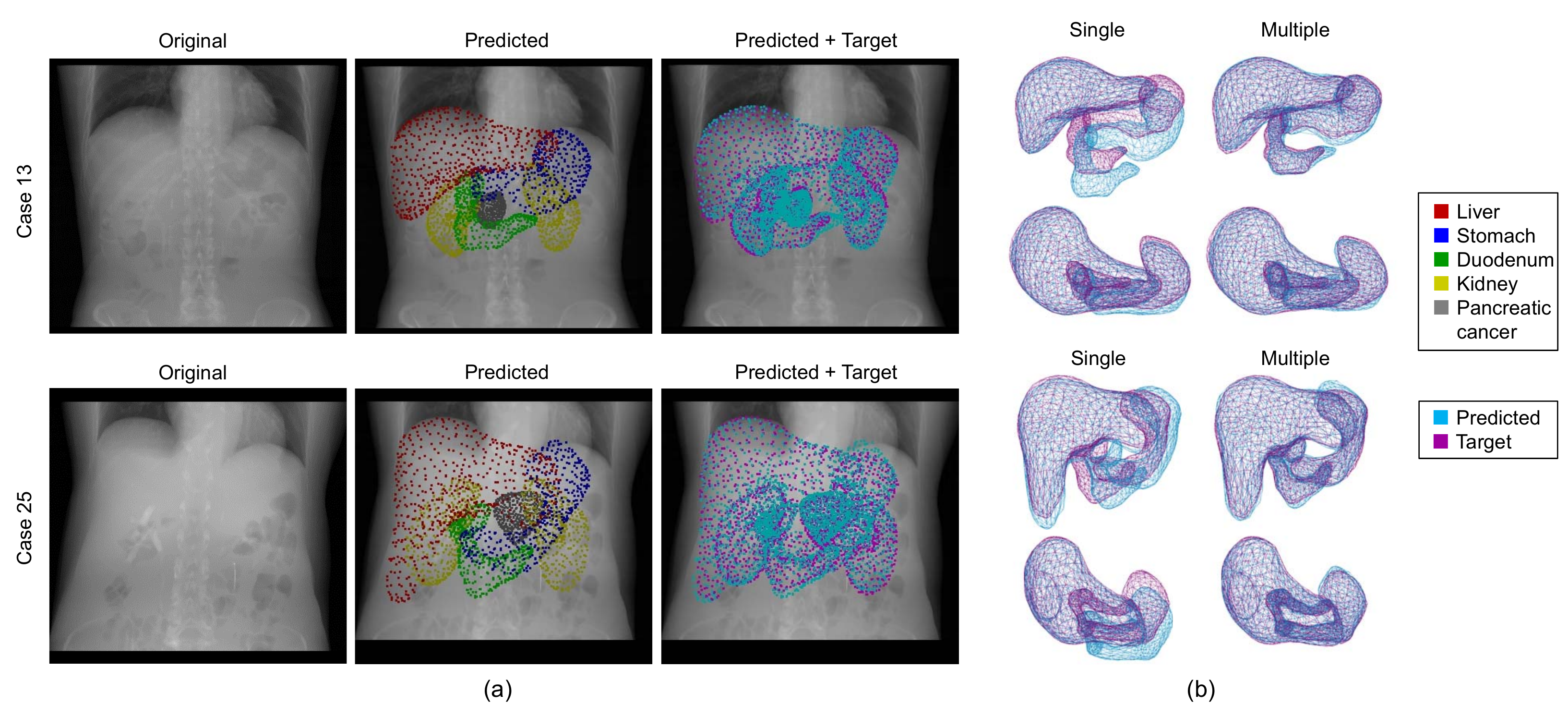}
	\end{center}
	\caption{Shape reconstruction examples and method comparison. Average and maximum error cases. The graph convolutions embed per-vertex displacement vector between organs, which results in better estimation performance of the regions with no visual cues.}
	\label{fig:8}
\end{figure*}

\subsubsection{Abdominal-organ shape reconstruction}
Cases 13 and 25 are shown as typical shape reconstruction examples in Fig. \ref{fig:8}; these cases show the average and maximum shape errors, respectively, for $t$ = 50, which had the largest displacement. Fig. \ref{fig:8}(a) shows the MR results, and the central image (predicted) shows the vertices of the 3D organ mesh obtained for the input DRR image, where coloring and superimposed visualization were conducted for each organ. The image on the right was obtained by projecting the true (magenta) and predicted (cyan) shapes on the projected image; the shape errors of each organ could be locally confirmed. Unlike with the liver, the contours could not be visually confirmed on the DRR images for the stomach, duodenum, and GTV; however, shape reconstruction with minimal deviation from the ground-truth shape was achieved. The supplemental movie (available online at \href{https://github.com/meguminakao/IGCN}{https://github.com/meguminakao/IGCN}) demonstrates the results for 10-frame sequential images.

Fig. \ref{fig:8}(b) shows the results obtained by visualizing the 3D organ meshes of the liver, stomach, and duodenum using SR and MR from two different directions. For SR, large deviations in the stomach and duodenum positions were noted. Thus, shape reconstruction using only the image features obtained from the 2D area in the DRR image corresponding to each organ was difficult. For MR, good matching was found between the true and predicted shapes. Note that stomach shapes varied considerably between patients because of the stomach contents, and some deviations were observed.

\section{Discussion}
This study presents a new framework that integrates an image generative network and GCN, which achieves model-based deformable registration for 2D projected images. Unlike image-based 2D/3D registration, a mesh that explicitly defines the organ areas to be estimated can be output. A wide range of clinical applications are possible, such as localization of GTVs and OAR volumes in radiation therapy, and tumor position identification for endoscopic camera images during surgery.

In conventional CNN-based feature encoding, image features away from the template-mesh projection points are referenced; this problem was resolved through displacement map generation by the image generative network. Additionally, significantly improved estimation accuracy for the stomach, duodenum, and pancreatic cancer GTV were confirmed through simultaneous reconstruction of multiple organs, even when there were no visual cues in the projected images. This outcome is thought to be due to successful localization through convolution of the features of adjacent vertices in the GCN, while referring to the image features and positional relationships between different organs.


Our experiments had certain limitations. For example, performance evaluation was performed using only DRRs as projected images. Thus, accuracy evaluations for X-ray images measured during treatment are required. However, multiple studies have reported that DRR-based learning is effective for prediction from measured X-ray images\cite{Ying19, Kasten20}. The organ shape contained in the 3D-CT data used to generate the DRR could potentially be taken as the true value, yielding a quantitative and highly reliable performance comparison, although the estimation error would increase because of the differences between the X-ray and DRR images. However, this increase would be limited even when random noise were added to the initial placement of the template mesh, and robust estimates could be expected for different imaging conditions in clinical settings.

\section{Conclusion}
In this study, we proposed an image-to-graph convolutional network (IGCN) that achieves deformable model registration of a 3D organ model for a single-viewpoint 2D projection image. We verified that clinically acceptable registration accuracy was achieved with a mean distance of less than 2 mm. The proposed technique could be directly applied for localization of radiation targets and organ-at-risk volumes in radiation therapy, and it could also be applied to a wide range of image-guided interventions.

\onecolumn

\section*{Supplementary document}

\subsection*{A. Method Comparison}

\begin{figure}[h]
	\begin{center}
		\includegraphics[width=17.5cm]{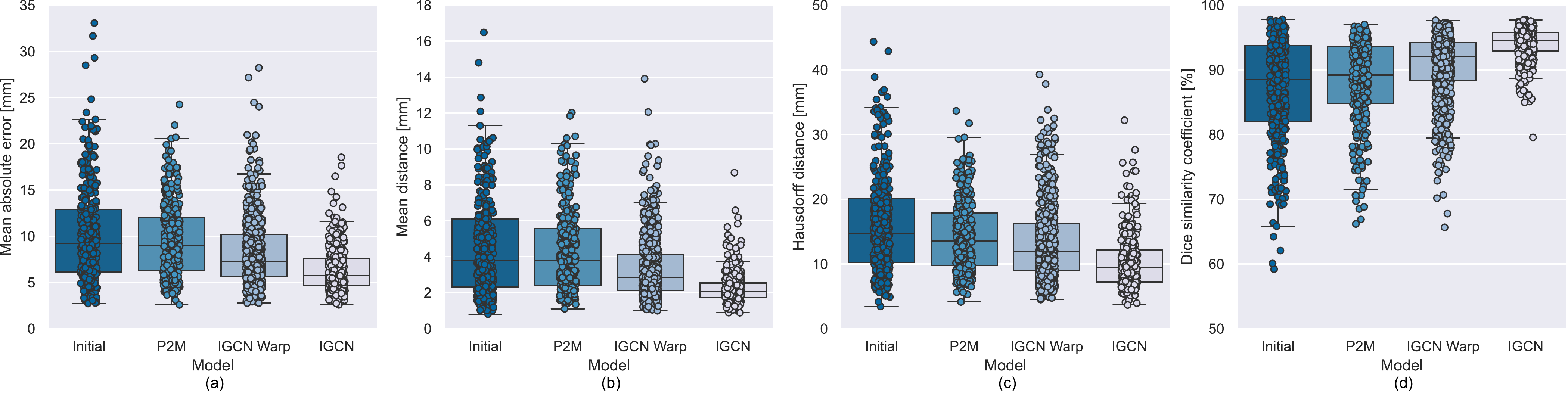}
	\end{center}
	\caption{Box and jitter plots of shape reconstruction results with respect to 350 liver shapes when noise was added to the initial template alignment: (a) MAEs, (b) MDs, (c) HDs and (d) DSCs.}
	\label{fig:8}
\end{figure}

\subsection*{B. Effectiveness of Statistical Generative Model}
We determined the effects of the different data augmentation methods on 3D organ shape reconstruction considering respiratory displacement variation. First, to determine the $(\omega_0, \omega_1, \omega_2)$ of the statistical generative model, training was conducted using eight parameter sets for 3D-CT models of 124 cases while generating variations in respiratory motion. The test data and subject organs were the same as in the section IV-B. We predicted the shape of the liver included in the 4D-CT data and calculated the MAE.

Table \ref{table:4} presents the investigated parameter set in order and obtained MAEs. When only $\omega_0$ was changed, relatively good performance was obtained for $\omega_0 = 2$, which corresponded to double the mean respiratory displacement. When both $\omega_0$ and $\omega_1$ were changed, improved (poorer) performance was obtained when the first (second) principal component was considered. Thus, we adopted $(\omega_0, \omega_1, \omega_2) = (2, 1, 0)$, for which the best performance was found. 

\begin{table}[h]
    \caption{Statistical augmentation results for different weight parameter sets}
	\label{table:4}

    \centering
    \scalebox{0.9}{
        \begin{tabular}{cccc}
    		\hline
    	    $\omega_0$ & $\omega_1$ & $\omega_2$ & MAE [mm]\\ 
    		\hline
    	    1 & 0 & 0 & 4.13 $\pm$ 0.14 \\
    	    2 & 0 & 0 & 3.99 $\pm$ 0.36 \\
    	    3 & 0 & 0 & 4.06 $\pm$ 0.41 \\
    	    1 & 1 & 0 & 4.01 $\pm$ 0.26 \\
    	    2 & 1 & 0 & 3.85 $\pm$ 0.22 \\
    	    2 & 2 & 0 & 4.49 $\pm$ 0.46 \\
    	    1 & 1 & 1 & 4.57 $\pm$ 0.31 \\
    	    2 & 1 & 1 & 4.05 $\pm$ 0.20 \\
    		\hline
        \end{tabular}
    }
\end{table}

Next, we compared the following cases: no data augmentation (``No augmentation''); application of random translation, which has traditionally been used as a data augmentation standard (``Random''); and data augmentation using the proposed statistical generative model (``Statistical''). For “Random,” twice the mean respiratory displacement was randomly applied in 3D regardless of direction; for ``Statistical,'' learning was conducted for a direction-dependent displacement based on the displacement vector corresponding to the average respiratory displacement vector and the first principal component, for weights obtained as described above.
Table \ref{table:5} lists the liver shape reconstruction errors for each data augmentation method. For each index, the best performance was achieved for learning using the statistical generative model, and the MAE decreased by 12.4$\%$ and 6.1$\%$ compared with ``No Augmentation'' and ``Random,'' respectively.

\begin{table}[h]
	\caption{Liver shape reconstruction results of augmentation methods}
	\label{table:5}

    \centering
    \scalebox{0.9}{
        \begin{tabular}{cccc}
    		\hline
    		\multirow{2}{*}{} & \multicolumn{3}{c}{Methods} \\ 
    		\cline{2-4} 
    	    & No augmentation & Random & Statistical (proposed)\\ 
    		\hline \hline
    	    MD [mm] & 2.51 $\pm$ 1.24 & 2.32 $\pm$ 1.00 & 2.10 $\pm$ 0.83 
    	    \\
    		HD [mm] & 11.34 $\pm$ 4.70 & 10.66 $\pm$ 4.18 & 9.90 $\pm$ 4.05 
    		\\
    		MAE [mm] & 6.94 $\pm$ 3.15 & 6.47 $\pm$ 2.48 & 6.08 $\pm$ 2.40
    		\\
    		DSC [\%] & 93.41 $\pm$ 3.24 & 93.91 $\pm$ 2.62 & 94.53 $\pm$ 2.24 
    		\\
    		\hline
        \end{tabular}
    }

\end{table}

\newpage 

\subsection*{C. Multi-organ shape reconstruction results}

\begin{figure}[h]
    \centering
	\includegraphics[width=175mm]{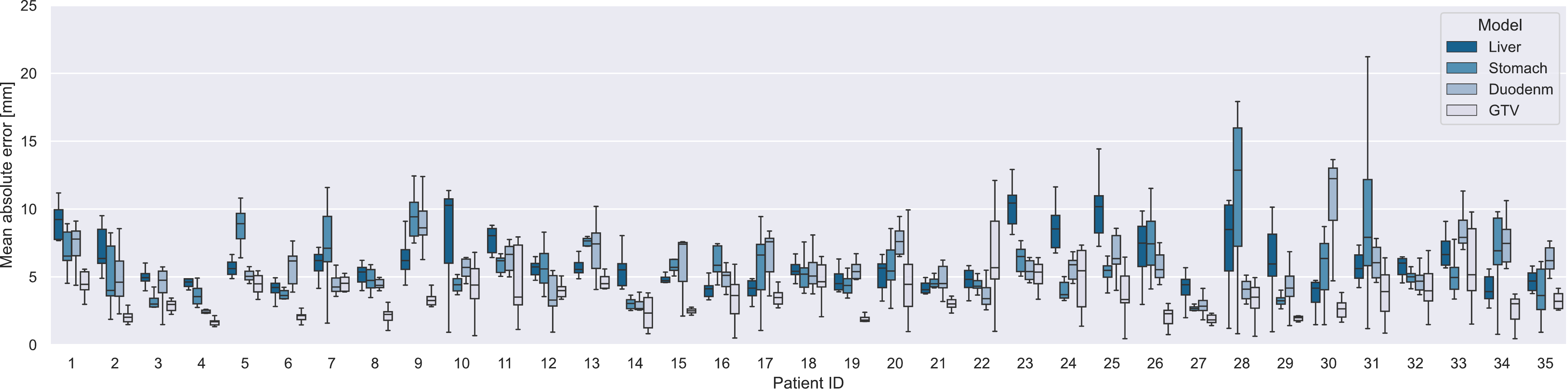}
	\caption{Shape reconstruction performance of multiple abdominal organs for all patient data. Each box plot represents MAEs for 10-frame images.}
	\label{fig:9}
	
	\vspace{1cm}
	
	\includegraphics[width=140mm]{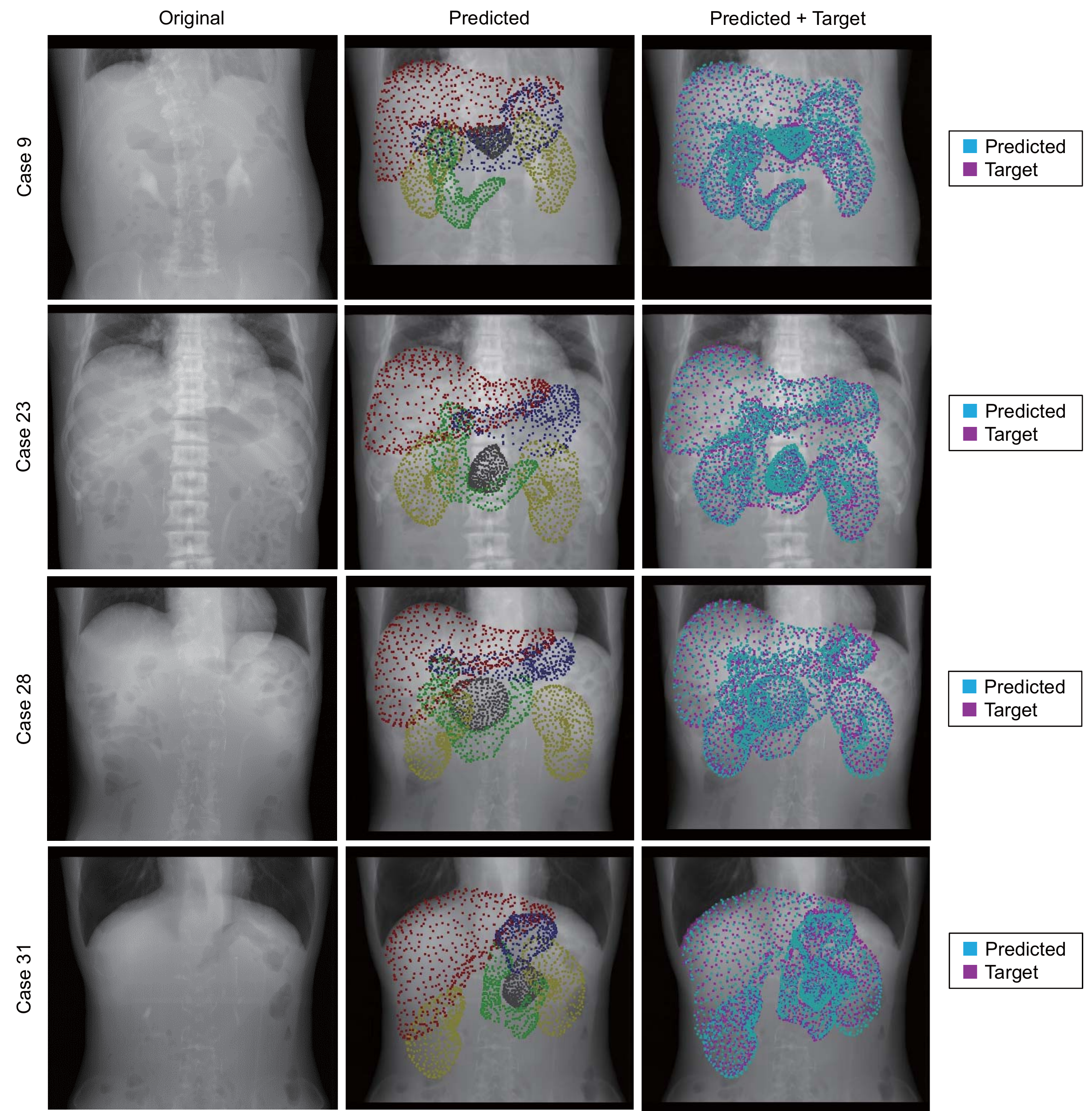}
	\caption{Multi-organ shape reconstruction results with large errors confirmed in liver and stomach. Despite that very low-contrast textures and different appearance of projection images are confirmed between patients, the shapes can be stably reconstructed with acceptable errors.}
	\label{fig:10}
\end{figure}

\end{document}